# Bandgap in orthorhombic thin lead halide perovskite (CsPbBr$_3$) nanosheets measured from STEM-EELS

Rosaria Brescia,[1] Stefano Toso,[2,3] Quentin Ramasse,[4,5] Liberato Manna,[2] Clive Downing,[6,7] Arrigo Calzolari,[8] Giovanni Bertoni[8*]

[1] Electron Microscopy Facility, Istituto Italiano di Tecnologia, Via Morego 30, 16163 Genova, Italy

[2] Nanochemistry Department, Istituto Italiano di Tecnologia, Via Morego 30, 16163 Genova, Italy

[3] International Doctoral Program in Science, Università Cattolica del Sacro Cuore, 25121 Brescia, Italy

[4] SuperSTEM, SciTech Daresbury Science and Innovation Campus, Keckwick Lane, Daresbury WA4 4AD, UK

[5] School of Chemical and Process Engineering & School of Physics, University of Leeds, Leeds LS29JT, UK

[6] CRANN and AMBER Research Centers, Trinity College Dublin, Dublin 2, Ireland

[7] School of Chemistry, Trinity College Dublin, Dublin 2, Ireland

[8] CNR - Istituto Nanoscienze, Via Campi 213/A, 41125 Modena, Italy

*Corresponding author: giovanni.bertoni@nano.cnr.it

## Abstract

Inorganic lead halide perovskites are promising candidates for optoelectronic applications, due to their bandgap tunability, high photoluminescence quantum yield, and narrow emission line widths. In particular, they offer the possibility to vary the bandgap as a function of the halide composition and dimension/shape of the crystals at the nanoscale. Here we present an aberration-corrected scanning transmission microscopy (STEM) study of extended nanosheets of $CsPbBr_3$ directly demonstrating their orthorhombic crystal structure and their lateral termination with Cs-Br planes. The bandgaps from individual nanosheets are measured by monochromated electron energy loss spectroscopy (EELS). We find an increase of the bandgap starting at thicknesses below 10 nm, confirming the less dramatic effect of 1D confinement in nanosheets compared to the 3D confinement observed in quantum dots, as predicted by density functional theory calculations and optical spectroscopy data from ensemble measurements.

## 1. Introduction

The interest towards lead halide perovskites (LHP) was boosted recently after the demonstration of high efficiency of this class of materials in photovoltaic devices in their thin film form, with the LHPs acting as absorbing layers. The power conversion efficiency of LHP-based solar cells has risen to >25% in the last years.[1] Among them, inorganic LHPs with $CsPbX_3$ (X = Cl-, Br-, and I-) composition have attracted increasing interest in recent years.[2,3,4] Further support for this research direction has come from the possibility to synthesize $CsPbX_3$ LHPs by colloidal routes,[5,6] and the ability to fabricate controlled sizes and geometries ranging from small nanocrystals such as nanocubes or nanoplatelets, to large ultrathin nanosheets.[7,8] The structure and optical properties of these nanostructures have been studied in details for pseudo-cubic and thin sheet nanocrystals,[9,10,11] but only recently has the bandgap from single nanocrystals with cubic shape been measured by high resolution electron energy-loss spectroscopy (EELS) in a scanning transmission electron microscope (STEM).[12]

The structure of $CsPbBr_3$ is commonly described as orthorhombic in bulk form (Pnma or Pbnm depending on the orientation of the crystallographic axes), while it can also be cubic (Pm-3m) in the case of nanocrystals.[6,13] Bulk $CsPbBr_3$ has been recently studied at very high resolution by combining converged beam electron diffraction and ptychography in an aberration-corrected STEM, pointing to a tetragonal (I4/mcm) symmetry, with out-of-phase rotated octahedra along the [001] direction at room temperature (RT).[14] However, a similar out-of-phase rotation is visible also in the [110] orientation of the orthorhombic structure (according to the Pbnm description).[15] The latter was confirmed in the case of small nanoplatelets by X-ray total scattering experiments.[16]

In summary, different possible structures were reported for nanostructures of $CsPbBr_3$ with reduced dimensions. In this communication we address the structure of $CsPbBr_3$ in the form of thin nanosheets (thickness $t$ down to ∼ 4 nm). This was made possible by state-of-the-art STEM making use of the advantages from probe correction (spatial resolution ∼ 80 pm), in addition to the high stability of the holder stage of the dedicated STEM instrument (Nion, Inc.). Moreover, these extended thin nanocrystals are ideal candidates for true spatially 1D confined systems (in the $z$ direction) with respect to 3D confined quantum dots (QDs), to verify the bandgap widening with decreasing thickness directly from individual nanosheets *via* high energy resolution monochromated electron energy-loss spectroscopy (EELS).

## 2. Experimental and Theoretical Methods

### 2.1. Synthesis of $CsPbBr_3$ nanosheets

The $CsPbBr_3$ nanosheets were prepared by an adaptation of the method first reported by Shamsi et. al.[17] First, a $PbBr_2$ stock solution was prepared by dissolving 0.13 g of $PbBr_2$, 2.5 ml of oleic acid (OA) and 2.5 ml of oleylamine (OLAM) in 25 ml of octadecene (ODE). The mixture was heated to 120 °C under vacuum and continuous stirring until the solid was completely dissolved. Upon cooling to room temperature, it remained limpid and stable for months. Second, a Cs-oleate stock solution in OA was prepared by dissolving 0.032 g of $Cs_2CO_3$ in 10 ml of OA, following the same procedure as the other solution. Finally, 3 ml of the $PbBr_2$ stock solution were put in a 20 ml glass vial, and a further 12 ml of ODE were added, to reach a total volume of 15 ml. The vial was closed with a rubber-septum cap and heated to 150 °C. Once the temperature was stabilized, 585 µl of octanoic acid (OCTAC) and 265 µl of octylamine (OCTAM) were introduced by piercing the septum with a needle and subsequently injecting the chemicals through it with the help of a micropipette. The temperature was left to equilibrate again to 150°C, and the stirring speed was set to 400 rpm. Then, 1 ml of the Cs-oleate solution was swiftly injected, and the temperature was kept constant to 150 ± 2 °C for 5 min. After this time, the solution was quenched by dipping the vial in a room-temperature water bath.

For the TEM analyses, one aliquot of the crude reaction mixture was diluted in an identical volume of hexane, and one single drop of the so-prepared solution was drop casted on a holey-carbon-film-coated Cu TEM grid. To remove the remaining excess of organics, the grid was washed by laying it on a paper tissue and drop casting 5 – 10 drops of pure hexane, paying attention that each drop was completely dried before casting the following one.

**2.2. Transmission electron microscopy and spectroscopy**

Prior to sample deposition, the commercial carbon support grid was cleaned by baking in vacuum (< $1\times10^{-6}$ mbar) at 200 °C for 7 hours. After drop-casting of the NS suspension, an additional gentle outgassing bake in similar vacuum conditions for 8 hours (at 55 °C) was applied to the TEM grid, to minimize carbon contamination during the experiments. The presented data were collected only on nanosheets overhanging the holes of the carbon support films, to avoid the absorption contribution from the carbon supporting film on the image contrast and on the EEL spectra. Atomic resolution images of the nanosheets were acquired at 200 kV acceleration voltage in annular dark field (ADF) STEM imaging mode on a Nion UltraSTEM200 microscope, equipped with a probe aberration corrector (TCD, Dublin, Ireland). The convergence semi-angle was 27 mrad and the inner cutoff angle of the STEM detector 99 mrad. STEM image simulations were calculated by using the software Dr. Probe.[18] To take into account the probe finite size in the experiment, the result was convolved with a Gaussian peak of 100 pm FWHM width. The choice of 200 kV for imaging thicker nanosheets assured high signal to noise imaging, while reducing the dominant ionization damage thanks to the high voltage.[19]

EEL spectra, in the form of 3D (*x, y, E*) datasets, were acquired at 60 kV on a Nion UltraSTEM100MC 'Hermes' aberration-corrected STEM, equipped with a Gatan Enfinium ERS EEL spectrometer optimized with high stability power supplies (SuperSTEM, Daresbury, UK). This instrument is equipped with a Nion high-resolution ground-potential monochromator allowing to obtain in these experiments a full width at half maximum (FWHM) of the elastic peak (zero-loss, ZL, peak) < 20 meV. The convergence semi-angle was 33 mrad and the collection semi-angle of the spectrometer was 44 mrad. The acquisition at 60 kV acceleration voltage gives better energy resolution and reduces the retardation effects on the energy loss spectrum, that may complicate the determination of the bandgap and depends on thickness.[20,21] However, at low kV, the ionization damage (or radiolysis) is higher,[22] resulting in a loss of features in the EEL spectrum. We were therefore particularly careful in spreading the electron dose for the EELS experiments to ensure as reliable a gap determination as possible. The beam current was set < 5 pA, and the spectra were collected in the Dual EELS acquisition mode, allowing the acquisition of two quasi-simultaneous spectra on the CCD camera of the spectrometer: one spectrum (low-loss) contains the ZL peak acquired at short exposure (10 msec), while the other spectrum (high-loss) contains the region of the bandgap acquired at longer exposure (100 msec). The total acquisition time for the two spectra at one pixel was about 110 msec. The low-loss spectrum is used to carefully align the spectra at sub-pixel level in the energy dimension by fitting a Gaussian peak under the ZL peak. The result is applied to the corresponding high-loss spectrum. We assume no drift in energy during the fast drift tube change of the spectrometer from low-loss to high-loss region in the acquisition of a single pixel of the dataset. Summing up short-acquisition spectra collected in the dataset over a large sample area, rather than acquiring a single long-acquisition spectrum, allows to obtain a reasonable signal-to-noise ratio at a reduced electron dose per unit area, thus at a reduced sample damage. The dispersion of the spectrometer was fixed to 2 meV or 5 meV per channel. No background subtraction from the spectra was performed to avoid extrapolation errors.

Energy dispersive X-ray spectra (EDS) were acquired with a Bruker solid state detector with a windowless 100 mm$^2$ effective area and were used to verify the atomic fraction of Pb and Cs in the sample.

Atomic structures in the figures were plotted using the software VESTA.[23]

**2.3. Density functional calculations**

First principles calculations, based on Density Functional Theory (DFT), were performed by using the Quantum Espresso package, for thin CsPbBr$_3$ films of different thicknesses ($N$ = 1, 2, 4, 6, 8, and 16 layers). Each film is simulated trough periodically repeated slabs, terminated by Cs-Br atoms and separated by ~1nm of vacuum in the direction perpendicular to the film, in order to avoid spurious interactions among replicas. A bulk structure was also simulated for reference. All the structures were relaxed self consistently until the forces on each atom are smaller than 0.03 eV/Å. We used the Perdew-Burke-Ernzerhof (PBE)[24] approximation to the exchange correlation functional for the calculation of total energy and band structure. Non-bonding interactions, such as van der Waals terms, are accounted for by the semiempirical method proposed by Grimme.[25] The hybrid PBE0 functional is used to correct the DFT deficiencies in reproducing the bandgap, without further atomic relaxation on the optimized PBE structures. Bandgaps are evaluated from the difference between the top of the valence band and the bottom of the conduction band. The bandgap values reported for the nanosheets were shifted by the difference between the calculated DFT bulk value and the experimental bandgap found at large thickness. Atomic potentials are described by ultrasoft pseudopotentials as available in the SSSP library.[26] Semi-core *5s5p* and *5d* electrons are explicitly included as valence electrons for Cs and Pb atoms, respectively. Single particle wavefunctions (charge) are expanded in planewaves up to a kinetic energy cutoff of 60 Ry (600 Ry), respectively. To test the effect of vacancies, neutral cation ($V_{Pb0}$, $V_{Cs0}$) and anion vacancy ($V_{Br0}$) point defects were simulated by placing a single vacancy (i.e. removing a single atom) in a (2x2x2) bulk superstructure. The Brillouin zone is sampled with a (6x6) *k*-point grid in the case of 2D films, and with (3x3x2) *k*-grid in the case of defective 3D structures.

## 3. Results and discussion

### 3.1. STEM imaging

The typical outcome of these synthesis processes consists in a distribution of thin rectangular nanosheets (NSs), a few hundreds of nm to a few μm in lateral size. A high-resolution ADF-STEM image from a $CsPbBr_3$ NS is presented in Figure 1. The columns formed only by Cs (Z = 55) atoms and only by Br (Z = 35) atoms, as well as the mixed Br and Pb (Z = 82) columns, are easily recognizable if the image is qualitatively compared with [001] projections of cubic (ICSD #97852) and orthorhombic $CsPbBr_3$ (ICSD #97851) structures, respectively. A close inspection at Br columns reveals that the image fits the [001] projection of the orthorhombic phase, characterized by a rotation of the Br octahedra in the ($a, b$) plane followed by a small tilt along the $c$ direction (Figure 1b). We used the Pbmn subset (space group n.62) for the choice of the crystal axes ($a$ = 8.36 Å and $b$ = 8.52 Å as derived from the experiment). This result is in agreement with the structure reported for the nanoplatelets (NPs) case.[16] Direct measurement of the lattice parameters $c$ is not directly possible from ADF-STEM due to the extremely large aspect ratio of the nanosheets. Figure 1c shows a (2x2) cell detail from the experimental image without any filtering (raw). Only a skew correction was applied to compensate the small distortion of the STEM scan. Figure 1d shows the same (2x2) cells after an average of 100 regions of the experimental image having the same size. The average image exhibits a clear reduction of the noise with negligible loss in resolution. The simulated STEM image obtained from the model in Figure 1b, for large thickness ($N$ = 16), is presented in Figure 1e. Even if the atomic positions are well reproduced by the model, an inspection of the contrast on the atomic columns can give information on the actual chemical composition. Figure 1f, g shows the contrast profiles along the [010] (center horizontal line) and [110] diagonal line from the STEM image. Along these two directions the relative contrast between Cs, Br, and Br-Pb atomic columns can be compared quantitatively with simulations. Clearly, a full stoichiometric $CsPbBr_3$ composition does not fit with the experimental data (blue lines). A model considering an occupancy $f$ = 0.70(5) of the Pb atoms matches much better with the experiment (orange lines). The result is not surprising, considering the high diffusion rate of Pb in similar perovskites. The Pb vacancies are expected to determine intrinsic *p*-doping. A DFT-calculated density of states shows that this picture holds for cation neutral vacancies ($V_{Pb0}$, $V_{Cs0}$), while the opposite behavior (i.e. *n*-doping) is observed in case of anion neutral vacancies ($V_{Br0}$), in agreement with the low (high) electronegativity of Pb and Cs (Br) atoms. However, in all cases, no deep defect states appear in the host bandgap due to vacancies. The higher contrast from the columns containing Pb atoms, the heaviest ones, is easily recognizable in the ADF STEM images, allowing to address the atomic termination at the edges. Figure 2 shows an ADF STEM image from an edge of a thin NS. It is clear that the last row of atoms corresponds to Cs-Br planes, in agreement with the surface termination reported by Bertolotti et al.[16] These correspond to (110)$_{Cs-Br}$ planes in the Pbmn structure. This has to be related to the preferential passivation of the ligands to Cs rather than Pb. We chose as a consequence a termination of the simulated slab in the [001] direction with (002)$_{Cs-Br}$ planes. Moreover, this termination agrees with a Pb stochiometric content < 1 as found from the ADF-STEM image simulations and from EDS quantification (Pb/Cs = 0.79(1)).

**3.2. EELS bandgap determination**

A representative EEL spectrum from a single $CsPbBr_3$ nanosheet is presented in Figure 3a. The spectrum is obtained after 200 spectra are summed up from an inner region of the nanosheet. An optical gap is evident at around 2.4-2.5 eV. The bandgap position was inferred from the maximum in the first-derivative spectrum, after a Savistzy-Golay smooth with 10 points intervals was applied to reduce fluctuations due to noise, following a similar procedure as in ref. 12. Similar data were obtained from a series of NSs of different thicknesses, and the determined band gap values as a function of thickness are presented in Figure 3b. As can be inferred from these results, the bulk bandgap is around 2.44 eV, and a small, yet appreciable widening of the gap starts to be visible below 20 layers, due to the electronic confinement in the *z* direction. The thickness of the nanosheets was measured according to the $t/\lambda$ ratio obtained in the very same positions but using a 0.020 eV dispersion (to better extrapolate the spectrum at higher energy loss), using the Iakoubovskii formula.[27] The maximum in the first derivative was obtained with a Gaussian fit, and an error of ± 5 meV was taken as a standard deviation. For the thickness, a relative error of ±10% was considered. The experimentally measured bandgap points were fitted with a power function according to the equation of confinement for a particle of mass $m^*$ in a crystal by impenetrable barriers in the z direction:[28]

$$\Delta E = E'_g - E'_{g,bulk} = \frac{\hbar^2 \pi^2}{2m^*}\left(\frac{1}{t^a}\right) \qquad \text{eq. 1}$$

Where $m^*$ is the effective mass of the exciton and $t$ is the thickness of the nanosheet, i.e. the confinement region or quantum well width along z. In a pure parabolic approximation, and considering a 1D model of the potential well with infinite barriers, the coefficient $a$ = 2.[28] Note that the gap energy includes the term from the binding energy of the exciton $E_b$ ($E'_g \sim E_g - E_b$). A better fit is obtained by assuming a $1/t$ dependence ($a$ = 1) (goodness $R^2$ = 0.83) in the formula rather than $1/t^2$ ($a$ = 2) (goodness $R^2$ = 0.79), the latter being predicted by a simple 1D model of the potential well with infinite barriers and parabolic bands, and confirming our previous findings.[8] However, the thinnest nanosheet found in the experiments corresponds to $N$ = 7 layers in thickness, with an expected emission still in the green region of the visible spectrum ($\lambda \geq$ 500 nm). DFT simulations agree very well with the experimental data, after aligning the values found in the bulk. Indeed, a change in the slope of the curve is clearly visible around $N$ = 8 layers, with bandgap rapidly increasing at lower thickness. This reflects the generally larger bandgap found by calculations for the orthorhombic structure (at the $\Gamma$ point of the Brillouin zone) with respect to the cubic structure (at the R point).[29] In Figure 3b, the quantum dots from Bekenstein et al.[7] (and from Brennan et al.[30] are also shown for comparison. These data are always above the predicted curve by DFT for the nanosheets, suggesting a different behavior in the case of quantum dots (3D confinement) and the nanosheets (1D confinement). The data point (◊) from orthorhombic nanoplatelets (NPs) from Bertolotti et al.[16] is also shown In Figure 3b. It agrees well with the present nanosheets behavior, and this is related to the similar anisotropy of the crystals in the two studies, with only one dimension out of the three being confined. The predicted bandgap value from DFT for the single layer (2.93 eV) is slightly higher than the one previously calculated (2.84 eV) assuming a cubic structure.[8] The results from Figure 3b can be used to predict the bandgap values in $CsPbBr_3$ nanosheets as a function of the layer thickness.

### 3.3. Elliott fit for inter-band optical gap estimation

As can be inferred from eq. 2, the variable $E'_g$ is related to the position of the excitonic peak. Indeed, a peak is clearly visible by extracting the dielectric function $\varepsilon = \varepsilon_1 + i\varepsilon_2$ from the EELS data with a Kramers-Kronig (K-K) analysis.[31] Figure 4a shows the so calculated dielectric function for the thick nanosheets. From the dielectric function, the absorption spectrum can be calculated (Figure 4b). To extract the inter-band value, we have to decouple the exciton peak from the band onset in the curve. This can be done approximately with an Elliot model of the absorption edge,[32,33,34] which considers a sum of discrete excitonic peaks added to a continuous profile (corrected at finite $T$ with the Fermi distribution) for the inter-band transition, and both are convoluted with the spectrum broadening (excitation life-time and experimental resolution) according to the equation (repeated here from supplementary equation 9 in ref. 32):

$$\alpha(\hbar\omega) \propto \frac{\mu_{cv}^2}{\hbar\omega}\left[\sum_J \frac{2E_b}{J^3}\operatorname{sech}\left(\frac{\hbar\omega-E_j^b}{\Gamma}\right) + \int_{E_g}^{\infty}\operatorname{sech}\left(\frac{\hbar\omega-E}{\Gamma}\right)\frac{1}{1-e^{-2\pi\sqrt{E_b/E-E_g}}}\frac{1}{1-\frac{8\mu b}{\hbar^3}(E-E_g)}\right] \quad \text{eq. 2}$$

The interested reader can find in ref. 32 a description of all the parameters. The model can be fitted to the spectrum to extract both $E_g$ (inter $-$ band gap) and $E_b$ (exciton binding energy). By applying this procedure to the bulk values (i.e., at large thickness) we obtain an offset of approximately 80 meV (see Figure 4b), which gives an estimation for the exciton binding energy.

## 4. Conclusions

STEM images at high resolution have enabled us to address the structure of CsPbBr$_3$ thin nanosheets. In particular, the Br atoms present the rotated octahedral disposition of the orthorhombic structure.[14] No evidence of the out of phase rotation of the Br octahedra that would be expected in the tetragonal symmetry was found, thus confirming our hypothesis of a predominantly orthorhombic structure. These thin and large nanosheets are ideal candidates to verify the emergence of a true 1D electronic confinement (in the $z$ direction). 60 kV EELS at very high resolution (FWHM < 20 meV) allowed to follow the bandgap opening as the CsPbBr$_3$ nanosheets became thinner. The extrapolation of the fit to extreme small thickness (single unit cell or $N$ = 2 layers) shows an overall lower confinement effect with respect to quantum dots and nanoplatelets (3D confinement) reported in literature. The effect of confinement starts to manifest itself at thicknesses below 10 nm, with a flex point between a low confinement regime and a high confinement regime. A bandgap of 2.446(5) at large thickness (bulk) was measured. This corresponds to 2.491(5) eV for the inter-band separation, according to Elliott theory. Monochromated EELS at low acceleration energy allows to precisely measures the bandgap in single nanocrystals semiconductors.

## Acknowledgements

SuperSTEM is the UK National Research Facility for Advanced Electron Microscopy, supported by the Engineering and Physical Research Council (EPSRC). R.B. and G.B. thank Valeria Nicolosi (Advanced Microscopy Laboratory, Trinity College Dublin) for the access to the microscopes. The authors thank Javad Shamsi of the University of Cambridge for the help in the synthesis.

# Figures

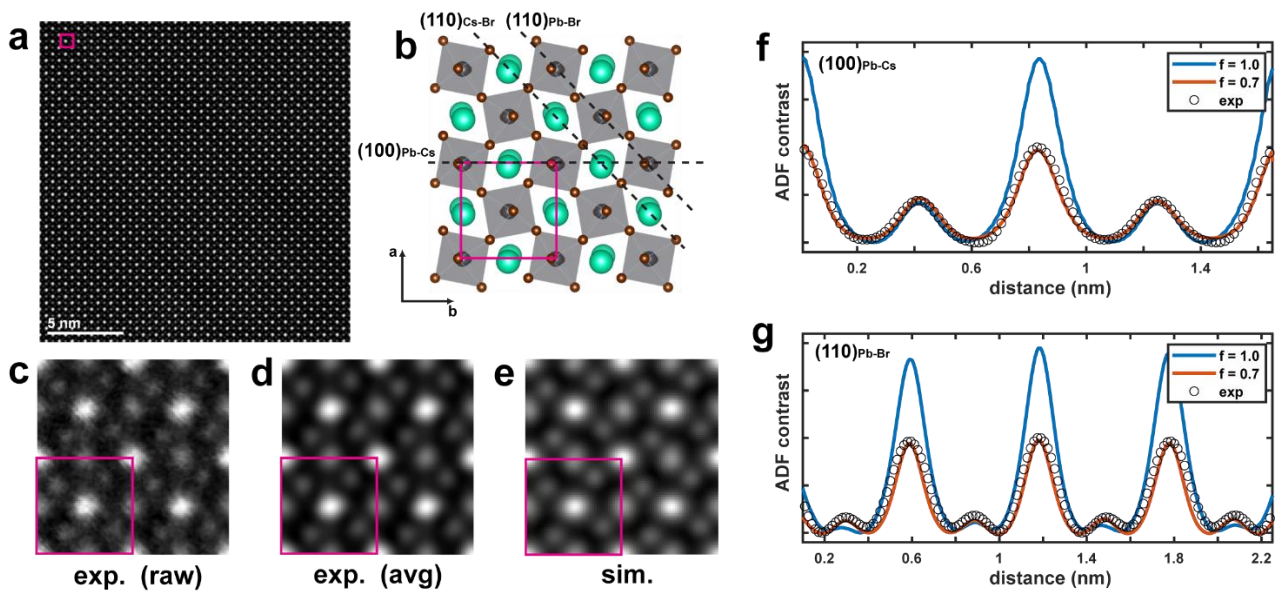

**Figure 1.** a) ADF-STEM image at 200 kV of an inner region of a CsPbBr$_3$ nanosheet in [001] orientation. The rotation of the Br octahedra in the (*a, b*) plane is clearly visible, as shown in the structure model in (b) (according to the Pbmn subset). In the model, the Cs atoms are in green, Br atoms in brown, and Pb are in the center of the grey octahedra. (Note that the actual edges of the crystals are at 45° with respect to *a* and *b*, see Figure 2. c,d) (2x2) cells region from the image before (raw) and after (avg) averaging 100 replicas from the same image. e) Simulation of the ADF-STEM image (2x2 cells) from the model in (b). f, g) Intensity profiles along the (100)$_{Pb-Cs}$ and (110)$_{Pb-Br}$ plane of the (2x2) structure in (d, e). A full stoichiometric CsPbBr$_3$ (full blue line) is compared with a sub-stoichiometric CsPb$_{0.7}$Br$_3$ (full orange line).

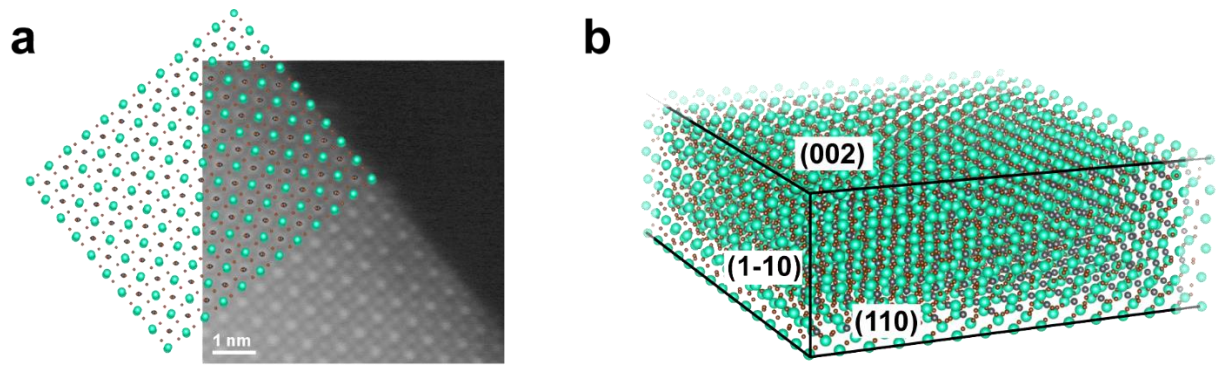

**Figure 2.** a) ADF-STEM image at 60 kV from the edge of a CsPbBr$_3$ nanosheet in [001] orientation. The Pb-containing atomic columns are easily recognizable by their higher contrast. The Pbmn structure from Figure 1 is superimposed for clarity (Pb is dark grey, Cs is green, and Br is brown). Clearly, the edge of the crystal is identified with Cs and Br atomic columns. These corresponds to (110)$_{Cs-Br}$ planes in the Pbmn structure of Figure 1. b) Sketch of the facets of the nanosheets as derived by ADF-STEM. The area of the (110) and (1-10) facets with respect to (002) is exaggerated for clarity.

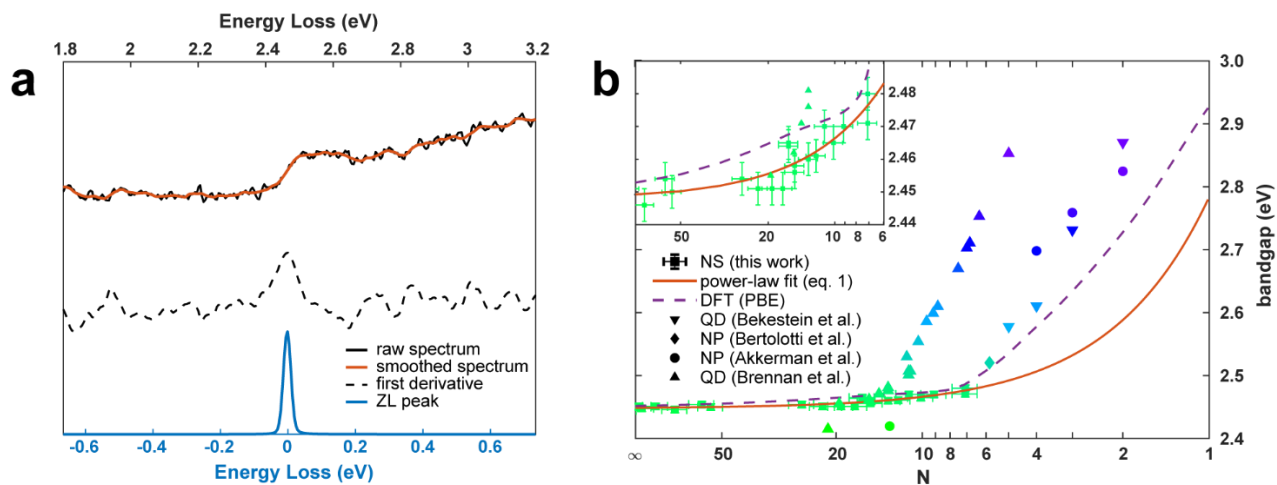

**Figure 3.** (a) EELS spectrum from a CsPbBr$_3$ nanosheet ($N \sim 50$) in the region of the optical bandgap. The black curve is the raw data, and the orange curve is the results after a Savitzky-Golay smooth. The dash black curve is the first derivative of the smoothed curve. The zero-loss (ZL) scaled spectrum used for the precise alignment of the spectra before summation is shown for comparison (blue). (b) Bandgap values calculated from the EELS

spectra as a function of the thickness (number of layers *N*) of the CsPbBr$_3$ nanosheets. The measured values are in green. The fit according to the power law function from eq. 1 is shown (orange line) a function of the number of layers *N*. The curve obtained from DFT simulations is shown in violet, together with values from CsPbBr$_3$ nanostructures as derived from literature. The inset shows an enlargement of the *N* > 6 layers region.

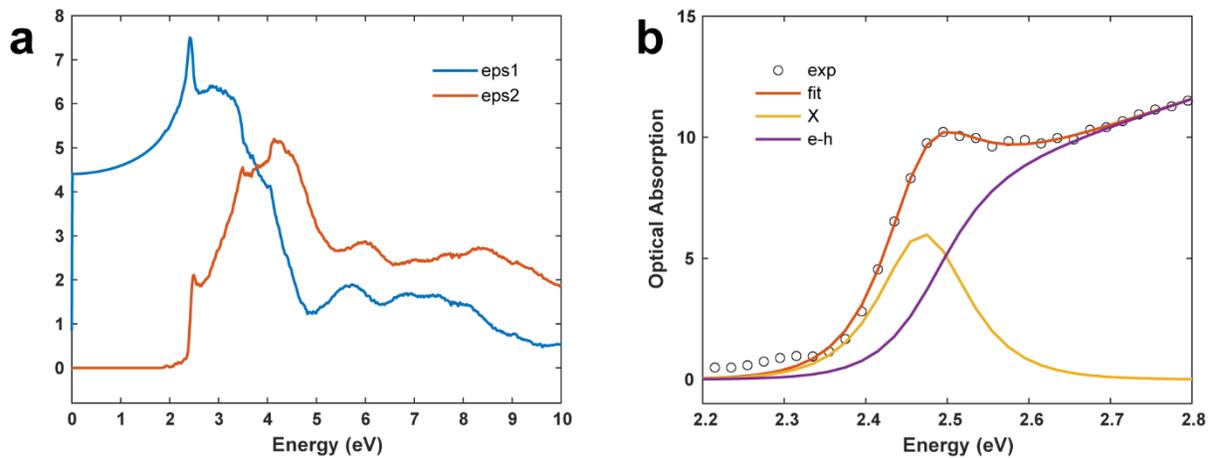

**Figure 4.** (a) Dielectric function of bulk CsPbBr$_3$ calculated from the EELS spectrum of a thick nanosheet (*N* ~ 50) according to the K-K theory, and assuming *n*(0) ~ 2.2.[35] (b) Enlarged view of the onset region of the optical absorption spectrum. The inter-band (*e-h*, purple curve) can be extracted from an Elliott fit. The exciton peak centered in $E_b$ (*X*, yellow curve) has a *sech* shape with broadening $\Gamma$ (FWHM). The offset with respect to the estimated values from the fit of Figure 2 is given by $E'_g - E_g \sim \frac{\Gamma}{2} + E_b$ ($E_b = 0.023$ eV, $E_g = 2.491$ eV, $\Gamma = 0.047$ eV).